\documentclass[11pt]{article}
\usepackage{amsmath,epsf,amssymb,latexsym,enumerate,cite,shadow,array,color}
\usepackage{slashed}
\usepackage{graphicx,wasysym}
 \pdfoutput=1

\DeclareFontFamily{U}{rsf}{} \DeclareFontShape{U}{rsf}{m}{n}{
  <5> <6> rsfs5 <7> <8> <9> rsfs7 <10-> rsfs10}{}
\DeclareMathAlphabet\Scr{U}{rsf}{m}{n} \makeatletter
\@addtoreset{equation}{section} \makeatother

\def\be{\begin{equation}}
\def\ee{\end{equation}}
\def\ba{\begin{array}}
\def\ea{\end{array}}

\newcommand{\bea}{\begin{eqnarray}}
\newcommand{\eea}{\end{eqnarray}}

\def\K{K{\"a}hler}
\def\vp{{\varphi}}

\newcommand{\ft}[2]{{\textstyle\frac{#1}{#2}}}

\usepackage{ifpdf}
\ifx\pdfoutput\undefined
   \pdffalse
\else
   \pdfoutput=1
   \pdftrue
  \usepackage[pdftex]{hyperref}
\newcommand{\rf}[1]{(\ref{#1})}

\begin{document}

\begin{titlepage}

\

\begin{center}
{\LARGE \textbf{Cosmological Attractor Models and Higher Curvature Supergravity 
\vskip 0.8cm }}

\

{\bf Sergio Cecotti${}^\dagger\,$\footnote{e-mail: {\tt cecotti@sissa.it}} and  Renata Kallosh ${}^\dagger\, {}^\dagger\,$\footnote{e-mail: {\tt kallosh@stanford.edu}}}  
\vskip 20pt

${}^\dagger$ Scuola Internazionale Superiore di Studi Avanzati, via Bonomea 265,  34100 Trieste, ITALY

${}^\dagger\, {}^\dagger\,$ Department of Physics and SITP, Stanford University, Stanford, California
94305 USA
\end{center}
\vskip 1.5 cm

\

\begin{abstract}

We study  cosmological $\alpha$-attractors  in superconformal/supergravity  models, where $\alpha$ is related to the geometry of the moduli space. 
For $\alpha=1$ attractors \cite{Kallosh:2013hoa}  we present   a generalization of the previously known manifestly superconformal
 higher curvature supergravity model \cite{Cecotti:1987sa}. The relevant standard 2-derivative supergravity with a minimum of two chiral multiplets is  shown to be dual  to  
a 4-derivative higher curvature supergravity, where in general  one of the chiral superfields  is traded for a curvature superfield. There is a degenerate case when both matter superfields become non-dynamical and there is only a chiral curvature superfield, pure higher derivative supergravity. Generic  $\alpha$-models \cite{Kallosh:2013yoa} interpolate between the attractor point at $\alpha=0$ and generic chaotic inflation models at large $\alpha$, in the limit when the inflaton moduli space becomes flat. They  have  higher derivative duals  with the same number of matter fields as the original theory or less, but at least one matter multiplet remains. In the context of these models, the detection of primordial gravity waves will provide  information on the curvature of the inflaton submanifold of the \K\, manifold,  and we will learn if the inflaton is a fundamental matter multiplet,  or can be replaced by a higher derivative curvature excitation.

 \end{abstract}

\vspace{24pt}
\end{titlepage}



\newpage

\section{Introduction}

The purpose of this paper is to study the dual relation between superconformal models underlying interesting cosmological models  which are in agreement with current  observations.
 In the past such a dual relation between the superconformal models/supergravities which have on one side  2-derivatives and scalars and on the other side higher curvature 4-derivative models, was established  in  \cite{Cecotti:1987sa}. More recently various aspects of such duality were clarified in \cite{Ferrara:2013rsa}, \cite{Ferrara:2013wka} in the context of cosmological models.

In this paper we will study two classes of cosmological models, $\alpha=1$ attractors \cite{Kallosh:2013hoa} and $\alpha\neq 1$ attractors \cite{Kallosh:2013yoa}. In the class of   $\alpha=1$ attractors,  on one side we have supersymmetric 2-derivative models with Einstein gravity and scalars, on the other side we have  supersymmetric models with 4-derivative  higher curvature  actions interacting with less scalars, in general. The other class of models,  $\alpha\neq 1$ attractors, may be divided in two subclasses. For the generic case we find that the 2-derivative \textsc{sugra} are not dual to 
local actions with higher curvature  actions and \emph{less} scalars. The duality transformation may be still performed, but the resulting $4$--derivative \textsc{sugra} has the \emph{same} number of scalars as the original theory.
There are, however, special $\alpha\neq 1$ models which have a $4$--derivative dual with one \emph{less} chiral field.

The bosonic  $\alpha=1$ attractors were introduced in  \cite{Kallosh:2013hoa}
 \begin{align}
  \mathcal{L}_{\alpha=1} = \sqrt{-g} \left[ {1\over 2} R - {1\over 2}  (\partial \varphi)^2 -   f^2\big(\tanh {\varphi\over\sqrt{6}}\big)
 \right] \, .
 \label{action}  \end{align} 
 where it was shown that the cosmological observables are universal and do not depend on the choice of the function $f$. They predict,  at large number of e-foldings inflation  $N$, which is between 50 and 60, the measurable tilt of the spectrum of fluctuations $n_s$ and the ratio of tensor to scalars fluctuations $r$ to be
$
 n_{s} = 1-2/N,   r= {12\over N^2}
$.
 A special choice of $f$  defines the Starobinsky model model \cite{Starobinsky:1980te}, \cite{Whitt:1984pd} with the potential
 \be
 f(\tanh)= M {2 \tanh\over 1+\tanh} \qquad \Rightarrow \qquad V= M^2  \big(1- e^{-\sqrt {2\over 3 } \varphi}\big)^2.
 \label{special}\ee
This model is dual to $R+R^2$ without scalars. The superconformal version of this model was presented in  \cite{Cecotti:1987sa} for the case corresponding to \rf{special}, where it was shown to correspond to a pure supergravity with 4-derivative interactions. 

In fact, attractors described in \rf {action} are not the most general ones, the ones in   \rf{action} give a simple example of functions with the following properties, as suggested in  \cite{Kallosh:2013hoa} 
\be
\tilde f(x) = a + b x^{-1} +\sum _{n=2} c_n x^{-n}
\label{genf}\ee
where $x= e^{\sqrt{2\over 3}\vp}$, so that $x^{-1}$ is small during inflation.

Here we will use the supersymmetric version of the models in \rf{action} and more general ones with  \rf{genf} constructed in \cite{Kallosh:2013hoa} and perform a duality transformation to a 4-derivative supergravity. We will see that not all scalars are removed in this duality transformation, in generic case. We will construct the relevant superconformal action with higher derivatives. A degenerate case of the duality relation when all dependence on scalars disappears, corresponding to  the choice \rf{special},  will be part of this generic model. In terms of \rf{genf} this choice is
 \be
\tilde f(x) = a + b x^{-1} 
\label{genfSt}\ee
The generalized bosonic  $\alpha$-attractor models  in  \cite{Kallosh:2013yoa}
 \begin{align}
  \mathcal{L}_{\alpha} = \sqrt{-g} \left[ {1\over 2} R - {1\over 2}  (\partial \varphi)^2 -   f^2\big(\tanh {\varphi\over\sqrt{6\alpha}}\big)
 \right] \, .
 \label{action1}  \end{align} 
predict at large $N$, for $\alpha $ not far from 1,
$
 n_{s} = 1-2/N$,   $r=\alpha \,  {12\over N^2}
$.
 At large $\alpha$, away from the attractor point,  the predictions are different, see eq.(5.2)-(5.4) in \cite{Kallosh:2013yoa}, they are not universal and depend on the choice of the function $f$. 
A special choice of $f(x)= M{2 x\over 1+x}$  defines the ${SU(1,1)\over U(1)}$ supersymmetric model  \cite{Ferrara:2013rsa} with the potential
 \be
 f(\tanh)= M {2 \tanh\over 1+\tanh} \qquad \Rightarrow \qquad V= M^2  \big(1- e^{-\sqrt {2\over 3 \alpha } \varphi}\big)^2
\label{fer} \ee
 The meaning of the parameter $\alpha$ was associated in \cite{Ferrara:2013rsa} with the \K\, manifold curvature, 
 \be
R_k= -{2\over 3 \alpha}
\label{curv}\ee 
This model continuously
interpolates between the prediction of the simplest chaotic inflationary model \cite{Linde:1983gd} with $V=\vp^2$  for
$\alpha \rightarrow \infty$, namely, $n_s= 1- {2\over N}$, $r= {8\over N}$, at large $N$,  the prediction of the Starobinsky model for $\alpha=1$, and the prediction
$n_s= 1-{2\over N}, \, r=0$ for $\alpha=0$,  as shown in Fig. 1. in \cite{Kallosh:2013yoa}.

In case of more general functions $f(\tanh)= (\tanh)^{n}$ in \rf{action1}
the models
 interpolate between the attractor point at $\alpha=0$  and generic chaotic inflation
models $\vp^{2n}$  \cite{Linde:1983gd} at large $\alpha$, as shown in Fig. 2 in \cite{Kallosh:2013yoa}. The corresponding prediction at large $N$ with $\alpha \rightarrow \infty$  is given by $n_s= 1- {n+1\over N}$ and $r= {8n\over N}$. For these models, as we will explain below, the holomorphic sectional curvature in the inflaton direction is given by $-{2\over 3 \alpha}$, and it vanishes at $\alpha \rightarrow \infty$.

Since both $r$ and $n_s$ are observables which will be measured with increasing precision during the next couple of decades, we find that in this class of models the curvature of the \K\, manifold \rf{curv} in \cite{Ferrara:2013rsa} (and coinciding with its  holomorphic sectional curvature in the inflaton direction in models in \cite{Kallosh:2013yoa}) is a cosmological  observable. In particular, if a detection of primordial gravity waves will take place and the value of $r$ will be known with high precision data on $n_s$, one might be able to evaluate $\alpha$ or find some constraints on it. 

The purpose of this note is to find a 
  dual relation between the supersymmetric version of the 2-derivative models with Einstein gravity interacting with 2 chiral multiplets  and the higher curvature models with less scalars. These two matter multiplets form a  minimal set, one is the inflaton, and the other is  a goldstino multiplet, required for stabilization.  In $\alpha=1$ case we find a simple duality which trades, generically, one of the chiral multiplets by a chiral curvature superfield. In particular case the model is just pure higher derivative supergravity. In  $\alpha\neq 1$ models 
\cite{Kallosh:2013yoa} duality to higher curvature models does  not reduce the original number of chiral matter multiplets, in a modified version of it, to be discussed below, 
 duality allows to remove one of the chiral matter superfields, which corresponds to a scalar coupled curvature. We find that all these  models do not reduce to pure higher curvature supergravity.

\section{Superconformal $\alpha=1$ attractors}

The superconformal action in general is defined by an embedding \K\, potential ${\cal N}$ and superpotential ${\cal W}$
\be
{1\over \sqrt{-g}}\mathcal{L}_{\rm sc}^{\rm scalar-grav}=-\ft16{\cal N} (X,\bar X)R
-G_{I\bar J}{\cal D}^\mu X^I\,{\cal D}_\mu \bar X^{\bar J}-G^{I\bar J}{\cal W}_I \bar{{\cal W}}_{\bar J} \, , \qquad I, \bar I = 0,1,2.
 \label{sc}
\end{equation}
Here $X^0$ is a conformon, $X^1 $ is an inflaton and $X^2$ is a goldstino multiplet, see for example \cite{Kallosh:2014ona} where this setting for our superconformal models is explained in details.

The  universality class  superconformal models are defined in \cite{Kallosh:2013hoa} as follows\footnote{Here we are using notation of  \cite{Kallosh:2013yoa} in application to the case $\alpha=1$.
}:
the embedding \K\, potential for these models has an $SU(1,1)$ symmetry between the complex conformon $X^0$ and the complex inflaton $X^1$ superfields.
\be
\mathcal{N}(X,\bar X)= -|X^0|^2 + |X^1|^2 + |X^2|^2  - 3 g\,  {(X^2\bar X^2)^2\over |X^0|^2-|X^1|^2 }\,  \,.
\label{calNminimal2}
 \ee
 The superpotential  preserves a subgroup of $SU(1,1)$, namely an $SO(1,1)$, 
 \be
{\cal W} = \sqrt{3}\, X^2 \Big ((X^0)^2- (X^1)^2\Big ) f (X^1/X^0)  \ .
 \label{sup}\ee
Function  $f (X^1/X^0)$ is invariant under local conformal-$\mathbb{R}$-symmetry, but if it is not a constant, it breaks the $SO(1,1)$.
\subsection{ Weyl and $\mathbb{R}$-symmetry gauge with $Z$ variables}

This gauge was studied for the model above in  \cite{Kallosh:2013hoa}
\be
 X^0= \bar X^0= \sqrt 3\, , 
\ee
where we  used  the local conformal and $U(1)$ $\mathbb{R}$-symmetry to impose this condition.
The remaining independent variables are defined as
\be
Z\equiv  {X^1\over X^0} \qquad S= \sqrt 3 {X^2\over X^0}
\ee 
Both $Z$ and $S$ have  vanishing conformal weight, $w=0$, and we are making this choice to fit the definitions in earlier papers.

This  gauge  leads to a supergravity version of the superconformal model \footnote{ We absorb numerical factors in $W$ into the arbitrariness of a function $f$. Moreover, the values of observables $n_s$ and $r$  depend only $V'/V$ and $V''/V$.  The value of $V$ during inflation, for example $M^2 $ in \rf{action}, is defined by  the amplitude of perturbations.}
\be
K= -3 \ln \Big [ 1-  |Z|^2- { |S^2|\over 3} +  g  \, {(S\bar S)^2\over 1-|Z|^2 } \Big], \qquad  W =  S \Big (1- Z^2\Big ) f (Z) \ .
\label{K}\ee
 The analysis of cosmology of this model was performed in \cite{Kallosh:2013hoa}, the term quartic in $S$ was necessary for the stability of the model when the scalar sgoldstino vanishes at the minimum of the potential at $S=0$.  This also requires the superpotential in this class of models to be linear in sgoldstino $S$. It leads to an attractor point in the $n_s-r$ plot for cosmological observables, for sufficiently  arbitrary choice of the function $f(Z)$, up to $1/N$ corrections. The simplest examples include the 
  T-Model, $f(Z)=Z^2$, or higher power of $Z$ where $N$ is the number of e-foldings of inflation. All these models are in   agreement with  the  current data  which are well described by a single inflaton scalar. Here we have 4 scalars from 2 superfields, but 3 of them are heavy near the inflationary trajectory,  they quickly go to their minimal values,  stop evolving and do not affect the evolution of the universe.

\subsection{ Weyl and   $\mathbb{R}$-symmetry  triangular gauge with $T$ variables}

Recently in \cite{Ferrara:2012ui} in the context of N=4 local superconformal model we have introduced the triangular gauge using the following gauge conditions.
We  impose an   $SU(1,1)$ invariant gauge to fix a Weyl symmetry
 \be
 |X^0|^2 - |X^1|^2= 3
 \ee
and for the $\mathbb{R}$-symmetry gauge we take
\be
{\rm Im} \, (X^0-X^1) =0
\ee
Our choice was motivated by the triangular decomposition of the $SL(2,\mathbb{R})$ matrix, when one switch from the $SU(1,1)$ matrix  to the $SL(2,\mathbb{R})$ basis, as shown in \cite{Ferrara:2012ui}  where we used the variable $\tau = \chi +i e^{-2\varphi}$.
We can therefore define an  independent variable $T= -i\tau $  as
\be
T=   {1+Z\over 1-Z} \, , \qquad Z= {T-1\over T+1}
\label{T}\ee
which parametrizes the coset space ${SL(2,\mathbb{R})\over U(1)}$, and $T=e^{-2\varphi}-i \chi$. The difference between these two gauges of the superconformal action \rf{sc} corresponds to a map between a disc and a half-plane. Meanwhile, the boundary of the moduli space, in proximity of which the inflationary evolution takes place, according to \cite{Kallosh:2013hoa},  for real $T$ and $Z$ is at $Z=1$ and at $T^{- 1}= 0$. Therefore in \rf{genf} the choice 
$\tilde f(T) = a + b T^{-1} +\sum _{n=2} c_n T^{-n}$ shows that near the boundary the terms with $n \geq 2$ are irrelevant.

\section{From generic $\alpha=1$ attractors to Cecotti's-type  higher derivative model}

We start with the \K\, potential   and superpotential \rf{K}  with arbitrary function $f(Z)$ and compare it with the simplest model of Cecotti \cite{Cecotti:1987sa} which corresponds to a supersymmetric version of the $R+ R^2$ model. For this purpose we 
substitute $Z$ as a function of $T$ from \rf{T} into \rf{K}.  We take into account that 
\be
 1-  |Z|^2- { |S^2|\over 3} + {g\over 3}  \, {(S\bar S)^2\over 1-|Z|^2 } = {2(T+\bar T)\over (T+1) (\bar T+1)}- { |S^2|\over 3} + {g\over 3}  \, { (T+1) (\bar T+1)(S\bar S)^2\over 2(T+\bar T)})
 \label{K'}\ee
We perform a holomorphic change of variables
\be
S= \sqrt 6 \, { C\over T+1}
\label{SandC}\ee
and find, ignoring numerical factors in front of the superpotential 
\be W=  S \Big (1- Z^2\Big ) f (Z) \Rightarrow  C { T\over (T+1)^3}\;
   f \!\left({T-1\over T+1}\right) \ .
 \label{W'}\ee
This leads to
\begin{multline}
-3 \ln\! \Big [{2(T+\bar T - C\bar C)\over (T+1) (\bar T+1)}+ 6 g \, { (C\bar C)^2\over (T+1) (\bar T+1)(T+\bar T)})\Big ]=\\
= 3\ln (T+1) (\bar T+1)-3\ln\! \left(T+\bar T - \bar C C +3 g  \, { (C\bar C)^2\over (T+\bar T)}\right)
\end{multline}
 We use the \K\, symmetry in $K$ and $W$ so that the equivalent theory is
\be
K=-3\ln\! \left(T+\bar T - \bar C C +3  g  \, { (C\bar C)^2\over (T+\bar T)}\right), \qquad W=  C \, T \,  \tilde f (T)=  C F( T )
\label{attractorT}\ee
Here
\be
f(Z)=f\Big ( Z={T-1\over T+1}\Big ) = \tilde f (T),  \qquad 
F(T)\equiv  \,T\, \tilde f(T)
\label{tilde}\ee
When $f(Z) = {2 Z\over 1+Z}$ it means that $\tilde f(T)= f \Big (Z={T-1\over T+1}\Big)={T-1\over 2 T}$ and at $g=0$, in absence of a stabilization term, we recover Cecotti's model \cite{Cecotti:1987sa} with 
\be
K=-3\ln (T+\bar T - \bar C C) \qquad W=  C (T-1).
\label{simple}\ee
This is eq.(19) in  \cite{Cecotti:1987sa}
 which is a supersymmetric generalization of the $R+  R^2$ model. In case of arbitrary function $f(Z)$ we have a more general class, in which Cecotti's model  \cite{Cecotti:1987sa} is one particular case. This is in agreement with 
\cite{Kallosh:2013lkr} where it was shown that Starobinski model \cite{Starobinsky:1980te}, \cite{Whitt:1984pd}
is associated with the potential $\Big ({2 \tanh \over 1+\tanh }\Big)^2$, whereas a generic attractor model depends on $f(\tanh (\varphi/\sqrt{6}))$.  In \cite{Kallosh:2013lkr} a stabilization term was added to \rf{simple}  in the form $ (\bar C C)^2$ without the $T$ dependence. The one in   \rf{attractorT}  was used for more general models in \cite{Kallosh:2013hoa},  \cite{Kallosh:2013yoa}.  We will discuss this important difference below, in secs. 4.1, 4.2.

\subsection{Equivalence with higher derivative supergravity for general superpotential}\label{equivalence1}
 Here we consider the dual relation between the superconformal cosmological attractor model \rf{calNminimal2}, \rf{sup}  (and its supergravity version \rf{K}) and higher curvature model. An intermediate version of the supergravity, which corresponds to a map between a disc and a half-plane and leads to \rf{attractorT} is used to establish this duality. First we simplify the problem  neglecting the stabilization terms, 
 i.e. we assume $g=0$. The cosmological model will be unstable at $g=0$, will not lead to a successful inflation, therefore  later we will study the effect of stabilization terms on duality.

\noindent\textbf{Claim.} \textit{The model in  \rf{calNminimal2}, \rf{sup}   is equivalent to a higher curvature supergravity with less matter fields for $g=0$.}\smallskip

\textbf{Notes:} 1) `equivalent' means equivalence of the classical equations of motion to the full non--linear level. 2) the equivalent higher--curvature supergravity may be written in different ways, with different field contents, and here we present just an example.\medskip

\textsc{Proof.} We write the superconformal form of the model (without bothering with $O(1)$ normalization coefficients)
\begin{align}\label{model}
-\big[\bar X^0 X^0 [( T+\bar T )-\bar C C]\big]_D+\Big(\big[C F(T)(X^0)^3\big]_F+h.c.\Big)\, , 
\end{align}
 $X^0$ is a chiral superconformal compensator field of weight $w=1$ whereas $T$ and $C$ have $w=0$. 
We define the curvature superconformal  field as in \cite{Cecotti:1987sa}
\begin{equation}\label{curvature}
{\cal R}=(X^0)^{-1}\Sigma(\bar X^0).
\end{equation}
Then \eqref{model} may be rewritten as
\begin{equation}\label{model2}
\big[\bar X^0 X^0 \bar C C\big]_D+\Big ( \big[(X^0)^2(X^0 C F(T)-T{\cal R}) \big]_F+h.c.\Big).
\end{equation}
{\it Since $T$ does not enter in the $D$--terms, its equation of motion is algebraic}
\begin{equation}\label{eom1}
X^0 C\,F^\prime(T)={\cal R} 
\end{equation}
and may be solved for either $C$ or $T$; for the particular form in eqn.\eqref{model} it is simpler to solve for $C$ (but the other one is classically legitimate too!)
\begin{equation}\label{whatC}
X^0 C={\cal R} \big/F^\prime(T)
\end{equation}
Plugging \eqref{whatC} in \eqref{model2}
 we get
\begin{equation}\label{model3}
\big[\overline{h(Q)}\, h(Q)\,\bar {\cal R}\, {\cal R}\big]_D+\Big(\big[Q\,(X^0)^2\,{\cal R}\big]_F+h.c.\Big)
\end{equation}
where the zero weight chiral superfield $Q\equiv y(T)$ and the holomorphic function $h(y)$ are implicitly defined by
\begin{equation}\label{implicitly}
\left\{\begin{aligned}&Q\equiv y(T)= T-\frac{F(T)}{F^\prime(T)}\\
 &h(T)=\frac{1}{F^\prime(T)}
\end{aligned}\right.
\end{equation}
Using \eqref{curvature} the lagrangian \eqref{model3} becomes (up to $O(1)$ normalization factors which are not keeping)
\begin{equation}
-\Big[\big(Q+\bar Q\big)\bar X^0 X^0\Big]_D+\Big[\overline{h(Q)}\, h(Q)\,\bar {\cal R}\, {\cal R}\Big]_D
\end{equation}
which is a model with just one chiral superfield $Q$  instead of $2$ living on the standard hyperbolic line $SU(1,1)/U(1)\equiv SL(2,\mathbb{R})/SO(2)$ coupled with $R^2$ terms with a non--trivial field dependent coefficient $|h(Q)|^2$ \textit{and No superpotential}. Of course, this last fact is just an artifact of the simplicity of the initial model and of the particular duality with higher curvature supergravity here chosen for elegance sake. 

In terms of the original fields we find
\begin{equation}
-\Big[\Big(T - \frac{F(T)}{F^\prime(T)}+h.c\Big)\bar X^0 X^0\Big]_D+\Big[\frac{1}{\overline {F^\prime (T)} F^\prime (T)}\,\bar {\cal R}\, {\cal R}\Big]_D
\label{GenC}\end{equation}
It means that the bosonic $R$-curvature dependent  terms are of the following form : the term linear in curvature enters as in Jordan frame, interacting with scalars, from the first superfield expression in \rf{GenC}
\begin{equation}
\Big(T - \frac{F(T)}{F^\prime(T)}+h.c\Big)R 
\label{bosGenC}\end{equation}
The term quadratic in $R$ also interacts with scalars, as we see from the second superfield expression in \rf{GenC}
\begin{equation}
\frac{1}{\overline {F^\prime (T)} F^\prime (T)}\, R^2
\label{bos2GenC}\end{equation}
Only when $F(T)$ is a linear function of $T$, both couplings disappear, see next subsection for a detailed discussion.

\subsection{Relation with the 1987 paper  \cite{Cecotti:1987sa}}

In that paper the model with $F(T)=a T+b$, $a$, $b$ constants, was shown to be equivalent to \emph{pure} supergravity with $R^2$ couplings. In that case both chiral superfields, $T$ and $C$ were integrated out, leaving a local Lagrangian. This choice corresponds to $c_n=0$ for $n\geq 2$ in \rf{genf}, or equivalently, to $f(\tanh)= {2\tanh\over 1+\tanh}$ in \rf{action}
\be
F(T)=a T+b \qquad   \qquad \tilde f(T) = a+ bT^{-1}.
\ee

That result is recovered by specializing the more general one above. Indeed, above we assumed that the relation $y=y(T)$ was invertible, as it is (locally) for generic functions $F(x)$. However, for the special function of the old paper this invertibility fails. From eqn.\eqref{implicitly} we get
\begin{equation}
T=-b/a\equiv \text{a constant}
\end{equation}
so the would--be chiral field $Q$ is actually a coupling constant and not a dynamical field. Then one gets
\begin{equation}
2\,\mathrm{Re}\!\left(\frac{2b}{a}\right)\;\Big[\bar X^0 X^0\Big]_D+ \frac{1}{|a|^2}\;\Big[\bar RR\Big]_D,
\end{equation}
which is the result of the 1987 paper  \cite{Cecotti:1987sa}. In components it  means that there is a term $\mathrm{Re}\!\left(\frac{2b}{a}\right) R$ and a term $\frac{R^2}{|a|^2}$ are present in the action, so that it is reduced to  \cite{Starobinsky:1980te}, \cite{Whitt:1984pd}.
For generic $F(T)$, however, one may integrate out only one chiral superfield {\it while preserving locality of the Lagrangian}. One ends up with a model \rf{GenC} which in components has a curvature terms $R$ in Jordan frame  with non-trivial coupling to scalars shown in \rf{bosGenC} as well as the terms $R^2$ coupled to scalars, shown in \rf{bos2GenC}, it is no longer {\it pure} 4-derivative  supergravity.

\section{Including stabilization terms in the duality relation}

\subsection{$T$-independent  stabilization term \cite{Kallosh:2013lkr}}\label{Tindepp}
Start with a standard \textsc{sugra} with the following form (in superconformal notation)
\begin{equation}\label{moregeneral}
 -\Big[\bar X^0 X^0 \Big ( T+\bar T\, -k(C,\bar C \Big )\Big]_D+\Big(\Big[C\, F(T)\,(X^0)^3\Big]_F+h.c.\Big)
\end{equation}
where the notations are as before. $k(z,\bar z)$ is any \emph{real} function, and $F(x)$ any \textit{holomorphic} function. The tachyon--free model  is the particular choice
\begin{equation}
 k(z,\bar z)= z \bar z-\frac{\zeta}{3} (z\,\bar z)^2,\qquad F(x)=3M(x-1).
\label{k}\end{equation}
It corresponds to a choice of stabilization made in  \cite{Kallosh:2013lkr}
\be
K=-3\ln\!\Big(T+\bar T - \bar C C +{\zeta\over 3}  \, { (C\bar C)^2}\Big) \qquad W=  \lambda \, C \, (T-1) 
\label{attractorSS}\ee
By the same gymnastics as before, \eqref{moregeneral} is rewritten as
\begin{equation}
 \Big[k(C,\bar C)\,X^0\,\bar X^0\Big]_D+\left( \Big[(X^0)^2\,X^0 C\,F(T)\,-T{\cal R}\Big]_F+h.c.\right),
\label{dual}\end{equation}
and again $T$ does not enter in the kinetic (\textit{i.e.}\! $D$--) terms and is an `auxiliary'
field which may be eliminated by its `algebraic' equations of motion without spoiling locality.
The equations of motion of $T$ have the same form as before, eqns.\eqref{eom1}\eqref{whatC}, $X^0 C={\cal R} \big/ F^\prime(T)$ and are solved by the same field $Q=T-\frac{F(T)}{F^\prime(T)}$  as before. 

Finally, we get
\begin{equation}\label{wwwcv}
-\Big[\big(Q+\bar Q\big)\bar X^0 X^0\Big]_D+\left[k\Big(\overline{h(Q)}\,\bar {\cal R}/\bar X^0, h(Q)\,{\cal R}/X^0\Big)\bar X^0\, X^0\right]_D
\end{equation}
where $h(Q)$ is the same function as before (eqn.\eqref{implicitly}).

This is again a good \textsc{sugra} model with $R^2$ couplings, and in fact a model of the class \textit{(B)} of the 1987 paper.
Again, for $F(x)=ax+b$ the chiral superfield $Q$ becomes non dynamical (a coupling constant) and we recover the special case of eqn.(23a) of that paper with no $Z^\alpha$ fields. In the general case we may use the duality of that old paper to get a standard \textsc{sugra} model with three chiral fields $Q$, $T$ and $C$. But a linear combination of $Q$ and $T$ decouples and we get back the original model with two fields $T$ and $C$ we started with, eqn.\eqref{moregeneral}.\medskip

\noindent\textbf{Remark.} The duality discussed above for general superpotentials $(X^0)^3 CF(T)$ may be seen as a \emph{particular} case of eqns.(23a),(23b) of the 1987 paper where we take just one superfield $Z$ and choose
\begin{equation}
 \Psi(z,\bar z,w,\bar w)= z+\bar z-k(w,\bar w).
\end{equation}
This gives an alternative proof of the result.\medskip

Eqn.\eqref{wwwcv} is a local \textsc{sugra} with a higher curvature coupling of the form $\boldsymbol{\Phi}\,R^2$, where now $\boldsymbol{\Phi}$ is a non--trivial function of the scalars of the chiral superfield $Q$ \emph{and} of the auxiliary fields of the (Poincar\'e) graviton supermultiplet (which propagate degrees of freedom in the higher curvature theory).

\subsection{Stabilization with $T$-independent masses of extra scalars, \cite{Kallosh:2013hoa}, \cite{Kallosh:2013yoa}}\label{inTT}
For more general attractor models in \cite{Kallosh:2013hoa} and later in \cite{Kallosh:2013yoa}  the following stabilization terms were used
\be
K=-3\ln\! \left(T+\bar T - \bar C C +3 g  \, { (C\bar C)^2\over (T+\bar T)}\right), \qquad W=   C \, F(T ).
\label{attractorStab}\ee
In such case, the mass formula for all 4 scalars  is given by the following expression \cite{Kallosh:2013yoa} for $\alpha=1$
 \begin{align}
  m_{{\rm Re}(\Phi)}^2 & = \eta_\varphi V \,, \quad
  m_{{\rm Im}(\Phi)}^2  = \left( \frac{4}{3 } + 2 \epsilon_\varphi - \eta_\varphi \right) V \,, \quad
  m_C^2 = \left( \frac{12 g  -2}{3 } + \epsilon_\varphi \right) V \,,
 \end{align}
where $\epsilon_\vp$ and $\eta_\vp$ are the slow-roll parameters of the effective single-field model \eqref{action}. In order to achieve stability, up to slow-roll suppressed corrections, one needs $g > 1/6$. When we compare this class of stabilization with the previous one, for some values of $T$, we find that the mass of the $C$ field depends on $T$. 
This makes the stabilization with $ {g  (C\bar C)^2\over (T+\bar T)}$ preferable to the one with $ {\zeta  (C\bar C)^2}$ since the mass of the $C$ field depends on
\be
3 g=   (T+\bar T) {\zeta\over 3}
\ee
For constant $g$ there is a simple and a universal limit on $g$, which stabilizes any such model. When we take the choice with constant $\zeta$, the mass of $C$ becomes a function of $T$ which needs to be studied and may or may not have problems, depending on the choice of a model and the specific values of $T$.

In case of generic stabilization with $T$-independent masses of non-inflaton fields, the dependence on $T$ remains in the D-term in the equation which generalizes \rf{dual}
\begin{equation}
 \Big[k(C,\bar C, T, \bar T)\,X^0\,\bar X^0\Big]_D+\left(\Big[ (X^0)^2\big(X^0 C\,F(T)\,-T{\cal R}\big)\Big]_F+h.c.\right).
\label{dual1}\end{equation}
In this case  our earlier duality transformation needs another generalization. 
We start with the superconformal  model
\begin{equation}\label{tindep}
-\left[X^0\bar X^0\left(T+\bar T-\bar C C+3g\,\frac{(\bar CC)^2}{T+\bar T}\right)\right]_D+\left(\Big[(X^0)^3 C\,F(T)\Big]_F+h.c.\right).
\end{equation}
For $g\neq 0$ this model is no longer equivalent to a higher derivative supergravity with a \emph{local} Lagrangian (having at most 4--derivatives) and \emph{less} chiral superfields, as it was the case for $g=0$. However, the theory is still classically dual to $R^2$ \textsc{sugra} coupled to the \emph{same} number (\textit{i.e.}, in the simplest model, $2$) of chiral superfields. 

To see this, let us introduce two new weight zero superconformal chiral fields $S$ and $Z$ and rewrite
\eqref{tindep}
in the form
\begin{equation}\label{lagranSZ}
-\left[X^0\bar X^0\left(T+\bar T-\bar C C+3g\,\frac{\bar Z Z}{T+\bar T}\right)\right]_D+\left(\Big[(X^0)^3 \big(C\,F(T)+S(Z-C^2\big)\Big]_F+h.c.\right).
\end{equation}
Integrating away $S$ produces a functional delta--function enforcing $Z=C^2$, which gives back \eqref{tindep}.
Now the equations of motion of $T$ and $Z$ read
\begin{align}
&\text{e.o.m. }\ T\colon &&\Sigma(\bar X^0)-3g\, Z\; \Sigma\!\left(\frac{\bar X^0\bar Z}{(T+\bar T)^2}\right)=(X^0)^2\, C\,F^\prime(T)\\
&\text{e.o.m. }\ Z\colon &&3g\, \Sigma\!\left(\frac{\bar X^0\bar Z}{T+\bar T}\right)=(X^0)^2\, S.
\end{align}
They may be solved for $C$ and $S$, respectively,
\begin{align}\label{eomT}
&C= (X^0)^{-2}\,\frac{1}{F^\prime(T)}\,\Sigma\!\left(\bar X^0\left(1-3g\, \frac{\bar Z Z}{(T+\bar T)^2}\right)\right)\equiv
(X^0)^{-1}\, \frac{1}{F^\prime(T)}\, \widetilde{\mathcal{R}}\\
&S= 3g\,(X^0)^{-2}\, \Sigma\!\left(\frac{\bar X^0\bar Z}{T+\bar T}\right),\label{eomZ}
\end{align}
where the last equality in eqn.\eqref{eomT} is the definition of the (local) superconformal chiral superfield $\widetilde{\mathcal{R}}$ which is equal to the usual scalar curvature multiplet $\mathcal{R}$, eqn.\eqref{curvature}, plus field--dependent corrections proportional to $g$.
Plugging back the expressions \eqref{eomT},\eqref{eomZ} in in eqn.\eqref{lagranSZ}, we get the `dual' higher derivative \textsc{sugra} coupled to the two chiral superfields $T$ and $Z$. We have
\begin{equation}
\Big[(X^0)^3 C F(T)\Big]_F+h.c.=\left[\bar X^0X^0 \left(1-3g\,\frac{Z\bar Z}{(T+\bar T)^2}\right)\left(\frac{F(T)}{F^\prime(T)}+\frac{\overline{F(T)}}{\overline{F^\prime(T)}}\right)\right]_D
\end{equation}
\begin{equation}
\Big[(X^0)^3 SZ\Big]_F+h.c.= 6g \left[\bar X^0X^0 \frac{\bar Z Z}{T+\bar T}\right]_D
\end{equation}
\begin{equation}
\Big[(X^0)^3 SC^2\Big]_F=3g \left[\frac{\bar X^0\bar Z}{X^0(T+\bar T)(F^\prime(T))^2}\,\widetilde{\mathcal{R}}^2\right]_D.
\end{equation}

In conclusion, the dual Lagrangian may be written in the form $L=L_\text{ordinary}+L_\text{higher derivative}$ where
\begin{gather}
L_\text{ordinary}=-\left[\bar X^0X^0\left\{T+\bar T-3g \frac{\bar Z Z}{T+\bar T}+\left(1-3g \frac{\bar Z Z}{(T+\bar T)^2}\right)\left(\frac{F(T)}{F^\prime(T)}+\frac{\overline{F(T)}}{\overline{F^\prime(T)}}\right)
\right\}\right]_D\\
L_\text{higher derivative}=\left[\frac{\overline{\widetilde{\mathcal{R}}
}\widetilde{\mathcal{R}}}{F^\prime(T)\,\overline{F^\prime(T)}}-\frac{3g}{\bar X^0X^0(T+\bar T)}\left(\frac{(\bar X^0)^2 \bar Z \widetilde{\mathcal{R}}^2}{(F^\prime(T))^2}+
\frac{(X^0)^2 Z \overline{\widetilde{\mathcal{R}}}^2}{(\overline{F^\prime(T)})^2}\right)\right]_D.
\end{gather}
Note that all $Z$ depend terms cancel as $g\to 0$, and we remain with the previous result, see sec.\ref{equivalence1}.

The lesson we learn here is that switching to a dual version with higher derivatives supergravity coupled to a number of scalars $\leq$ the original one is still possible; however, in most cases the models are not getting simpler, with exception of the case in which we limit to deform the elegant model corresponding to pure higher derivative supergravity without scalars by $T$-independent stabilization terms as in sec.\;\ref{Tindepp}.

\section{Superconformal $\alpha$-attractors}

The supersymmetric $\alpha$-attractor  model was proposed in \cite{Kallosh:2013yoa}, following the first supersymmetric cosmological model in this class with $\alpha\neq 1$, which  was suggested in \cite{Ferrara:2013rsa}, see eq. \rf{fer} there. The superconformal model in \cite{Kallosh:2013yoa}, \cite{Kallosh:2014ona}, after the map of the disk to the half-plane as we described above, is
\begin{align}\label{alpha}
-\bigg[ \bar X^0 X^0 \left(T + \bar T   -  C \bar C  
 + 3 g \frac{( C \bar C)^2}{T + \bar T}\right)^{\!\!\alpha}\, \bigg]_D+\Big(\big[C F(T)(X^0)^3\big]_F+h.c.\Big)
\end{align}
which coincides with the previous one at $\alpha=1$. The model on the disk is given in eqs. (4.1), (4.2) in \cite{Kallosh:2013yoa} and it becomes more elegant when transformed to half-plane as we see in \rf{alpha}.
Since the K\"ahler metric is just rescaled by a factor $\alpha$, the scalar curvature $R_k$ of the \K\, manifold for these \textsc{sugra} models is rescaled by $1/\alpha$, and then at $C=0$ is 
\be\label{curv1}
R_k\Big|_{C=0}= - {2(1-2g)\over \alpha}.
\ee
However, since the field $C$ is massive, the relevant curvature for classifying universality classes of inflation is not the scalar curvature of the K\"ahler manifold $R_k$ but rather the holomorphic sectional curvature in the $\partial_T$ direction along the half--plane $C=0$ defined by the vev of the massive chiral multiplet $C$. For the models \eqref{alpha} the relevant sectional curvature is
\begin{equation}\label{rrrffds}
\text{holomorphic sectional curvature}\equiv \left.\frac{R_{T\bar TT\bar T}}{(G_{T\bar T})^2}\right|_{C=0}= -\frac{2}{3\alpha},
\end{equation}   
which agrees with the scalar curvature computed using the induced metric on the half--plane $C=0$, 
\begin{equation}
G_{T\bar T}^\text{ind}=-3\alpha\,\partial_T\partial_{\bar T}\ln(T+\bar T),
\end{equation}
since the half--plane $C=0$
 is a totally geodesic submanifold of the full K\"ahler space.
 Note that the sectional curvature is $g$ independent.
 This result is quite general: if our \textsc{sugra} model contains a number of chiral superfields $T$, $C_i$, ($i=1,\dots, m$) with K\"ahler potential $K(T,\bar T, C_i,\bar C_i)$ and \emph{all} the scalars in $C_i$ are massive, the K\"ahler submanifold $C_i=\langle C_i\rangle$, is totally geodesic and the sectional curvature along $\partial_T$, restricted on that submanifold, may be computed as the scalar curvature of the one--dimensional K\"ahler manifold
with induced K\"ahler potential
\begin{equation}
K^\mathrm{ind}(T,\bar T)= K\!\Big(T,\bar T, \langle C_i\rangle, \langle \bar C_i\rangle\Big).
\end{equation}
If, as in the models \eqref{alpha}, the induced metric has a $SU(1,1)$ symmetry, and hence is isometric to the standard Poincar\'e metric up to an overall rescaling by a factor $\alpha$, we get
eqn.\eqref{rrrffds} in full generality.

However, the above is not the only possible construction. Indeed, supersymmetry is broken and, in particular, the two scalars of the $T$ multiplet have different masses: $\mathrm{Im}\,T$ is much heavier than $\mathrm{Re}\,T$ and it makes sense to fix  to its vacuum value, $\mathrm{Im}\,T=0$. Then the submanifold of the K\"ahler space spanned by the light scalars is just a \emph{real} curve parametrized by $\mathrm{Re}\,T$ which happens to be a geodesic of the full K\"ahler geometry. In this \emph{real} sense the universality classes are defined by the geometry along the geodesic. Two models, with different K\"ahler metrics, which are isometric along the respective geodesics $\mathrm{Im}\,T=0$, $C_i=\langle C_i\rangle$ belong to the same universality class. Note that the sectional curvature along the geodesic --- which may still be defined --- is \emph{not} a universality class invariant any longer (since a one--dimensional space has no intrinsic curvature). The role of the curvature as an invariant is specific to the class of models with an effective $SU(1,1)$ isometry of the reduced K\"ahler space $C_i=\langle C_i\rangle$.   

An example 
 of a supersymmetric $\alpha$-attractor model, which is isometric to \eqref{alpha} only when restricted to the real curve
 $\mathrm{Im}\,T=0$, $C=0$, leading to the same inflationary cosmology action in \rf{action1}, is described in \cite{KL}. It corresponds to the superconformal model
\begin{align}\label{alpha3}
- \left[\bar X^0 X^0 \left(T + \bar T  + {\alpha-1 \over 2} {(T - \bar T)^2\over T+\bar T}  -  C \bar C  
 + 3 g \frac{( C \bar C)^2}{T + \bar T}\right)\right]_D+\Big(\big[C F(T)(X^0)^3\big]_F+h.c.\Big)
\end{align}
which coincides with the previous one at $\alpha=1$. 
Although, in this case, the curvatures are
not useful to characterize the universality class,
for the sake of comparison we list them (along the light scalar curve $T=\bar T$ and $C=0$) 
\begin{align}\label{curv2}
&\text{scalar curvature}\colon &R_k\Big|_{T=\bar T\atop C=0}&= - {2\over 3\alpha} +\alpha^{-2}- \frac{7}{3}+4g\\
&\text{sectional curvature}\colon &\left.\frac{R_{T\bar TT\bar T}}{(g_{T\bar T})^2}\right|_{T=\bar T\atop C=0}&= - {2\over 3\alpha} +\alpha^{-2}- 1\label{piiq}
\end{align}
Note that eqns.\eqref{curv1},\eqref{curv2} and eqns.\eqref{rrrffds}, \eqref{piiq} agree at $\alpha=1$. \medskip

For the purpose of duality, it is convenient to divide the $\alpha$--attractor models in two classes having different properties. The second class correspond to the generic
$\alpha$--attractor models, while the first one contains \emph{special} models which have better duality properties than the generic ones. We shall refer to the two classes as \emph{special} and \emph{generic} $\alpha$--attractor models, respectively. Note that at $\alpha\rightarrow \infty$ the sectional 
curvature of \emph{generic} $\alpha$--attractor models in \eqref{rrrffds} vanishes, but the one in  \eqref{piiq} for \emph{special} models has the value $-1$ in the limit.

\subsection{Special $\alpha$--attractor models}

The prototypical special $\alpha$--attractor model is the one
in eqn.\eqref{alpha3}
where we ignore the stabilization term  and take $g=0$.
However, general $T$--independent stabilization terms, as the ones considered in sec.\;\ref{Tindepp}, are still allowed. 

It is convenient to study the duality properties of a larger class of $N=1$ supergravities. Then
suppose we have a $N=1$ \textsc{sugra} whose Lagrangian has (in superconformal notation) one of the following \emph{two} forms:
\begin{multline}\label{lagrangian1}
I)\quad -\Big[\bar X^0X^0\Big(G(T,\bar T, Z_i,\bar Z_i)+ H(C,\bar C, Z_i, \bar Z_i)\Big)\Big]_D+\\
+\Big(\Big[(X^0)^3\Big(C F(T, Z_i)+L(T,Z_i)+V(C,Z_i)\Big) \Big]_F+h.c.\Big),
\end{multline}
\begin{multline}\label{lagrangian2}
II)\quad -\Big[\bar X^0X^0\Big(G(T,\bar T, Z_i,\bar Z_i)+ H(C,\bar C, Z_i, \bar Z_i)\Big)\Big]_D+\\
+\Big(\Big[(X^0)^3\Big(\log\!\big[C F(T,Z_i)+D(T,Z_i)\big] +L(T,Z_i)+V(C,Z_i)\Big) \Big]_F+h.c.\Big),
\end{multline}
where  $T, C, Z_i$ ($i=1,2\dots,n$) are $w=0$ superconformal chiral multiplets, and $G$, $H$, $F$, $L$, $V$, $D$ are arbitrary functions of their arguments ($G$, $H$ being real and the others holomorphic). For instance, we may consider the $I)$ model with $n=0$, $L=V=0$, and 
\begin{gather}\label{modelspecial}
G(T,\bar T, Z_i,\bar Z_i) = T+\bar T+\frac{\alpha-1}{2}\frac{(T-\bar T)^2}{T+\bar T}\\
H(C,\bar C,Z_i, \bar Z_i)= -\bar C \bar C +\frac{\zeta}{3} (C\bar C)^2
\end{gather}
which corresponds to the $\alpha$--attractor model \eqref{alpha3} with the stabilizing term of sec.\;\ref{Tindepp}. 
Other interesting models may be set in one of the forms $I),II)$ by an appropriate field redefinition.

The $T$ equations of motion read
\begin{align}
&I)\colon &&X^0\Sigma[\bar X^0\partial_T G]= (X^0)^3 \Big(C\, \partial_T F(T,Z_i)+\partial_TL(T,Z_i)\Big)\\
&II)\colon &&X^0\Sigma[\bar X^0\partial_T G]= (X^0)^3 \left(\frac{C\, \partial_T F(T,Z_i)+\partial_TD(T,Z_i)}{C F(T,Z_i)+D(T,Z_i)}+\partial_TL(T,Z_i)\right)
\end{align}
which have the property that they may be explicitly solved for $C$. In case $I)$ we get
\begin{equation}\label{thisequation}
C= (X^0)^{-2} \frac{1}{\partial_T F(T,Z_i)}\Sigma[\bar X^0\partial_T G]-\frac{\partial_T L(T,Z_i)}{\partial_T F(T,Z_i)}\equiv  \frac{(X^0)^{-1}\mathcal{R}_G}{\partial_T F(T,Z_i)}-\frac{\partial_T L(T,Z_i)}{\partial_T F(T,Z_i)},
\end{equation}
where the second equality is the definition of the higher curvature chiral superfield $\mathcal{R}_G$ (which reduces to the standard one for $G$ linear in $T$ as in the $\alpha=1$ models)
\begin{equation}\label{RGC}
\mathcal{R}_G=  (X^0)^{-1}\Sigma[\bar X^0\partial_T G].
\end{equation}
In case $II)$ eqn.\eqref{thisequation} get replaced by
\begin{equation}\label{thisequationII}
C= \frac{D\,(X^0)^{-1}\mathcal{R}_G+\partial_TD+D\partial_TL}{F\,(X^0)^{-1}\mathcal{R}_G -\partial_TF-F\partial_TL}.
\end{equation}
Replacing back eqn.\eqref{thisequation} (resp.\! eqn.\eqref{thisequationII}) into 
eqn.\eqref{lagrangian1} (resp.\! eqn.\eqref{lagrangian2}) we get a standard form higher derivative \textsc{sugra} with a local Lagrangian with one less chiral field (namely $C$ which was integrated away using the $T$ equations). Of course, the equivalence of the two theories is purely classical.

\textit{Thus \emph{special} $\alpha$--attractor models are dual to a 4--derivative \textsc{sugra} with one less chiral superfield.}

For instance, for $I)$ models with $n=L=V=0$, we get the Lagrangian
\begin{equation}\label{rrtyu}
-\left[\bar X^0X^0\left(G(T,\bar T)- \frac{F(T)}{F^\prime(T)}\partial_T G-\frac{\overline{F(T)}}{\overline{F^\prime(T)}}\partial_{\bar T}G\right)\right]_D+\left[H\!\!\left(\frac{(X^0)^{-1}\mathcal{R}_G}{F^\prime(T)}, \frac{(\bar X^0)^{-1}\overline{\mathcal{R}}_G}{\overline{F^\prime(T)}}\right)\right]_D,
\end{equation} 
so that the dual K\"ahler potential is
\begin{equation}
K=-3\ln\!\left(G(T,\bar T)- \frac{F(T)}{F^\prime(T)}\partial_T G-\frac{\overline{F(T)}}{\overline{F^\prime(T)}}\partial_{\bar T}G\right).
\end{equation}

If $F(T)=aT$ is a linear function and, as in eqn.\eqref{modelspecial}, $G(T,\bar T)$ is homogeneous of degree $1$ in $T,\bar T$ so that
$T\partial_T G+\bar T\partial_T G\equiv G$, the first term in
\eqref{rrtyu} vanishes, leaving only the higher curvature coupling.

This system reduces to pure higher derivative supergravity only if the function G is chosen so that the Kahler metric is isometric to $SU(1,2)/U(2)$ with standard ($\alpha=1$) curvature, as it is easy to see from the above expressions. In general we have ordinary higher derivative supergravity coupled to $T$ and the spectator fields $Z_i$, if any.
The Lagrangian \eqref{rrtyu} contains the generalized 
curvature chiral superfield $\mathcal{R}_G$, eqn.\eqref{RGC}. To gain some physical intuition on the 4--derivative supergravities with Lagrangians in the form \eqref{rrtyu},
we rewrite $\mathcal{R}_G$ in a more suggestive fashion. 
After, possibly, a redefinition $T\to T+\mathrm{const.}$ to make $T=0$ a point at which $G$ is analytic,
we may write with no loss of generality $G= \sum_{n,m} a_{n,m}\, T^n \bar T^m$, where the coefficients $a_{n,m}$ may depend on the chiral fields $Z_i$ but not on $T$ or $C$.
For simplicity, we assume no $Z_i $ field to be present. Then $\partial_T G= \sum_{n,m}  n\, a_{n,m}\, T^{n-1} \bar T^m$, and we may write
\begin{equation}
\mathcal{R}_G= \sum_{n,m} n\, a_{n,m}\, T^{n+m-1} (X^0 T^m)^{-1} \Sigma[\bar X^0 \bar T^m] = \sum_{n,m} n\, a_{n,m}\, T^{n+m-1} \mathcal{R}_m
\end{equation}
where we defined
\begin{equation}
\mathcal{R}_m= (X^0 T^m)^{-1}\, \Sigma[\bar X^0 \bar T^m].\end{equation}
Note that for a \emph{fixed} $m$ the redefinition $X^0 \to  X^0 T^m$ makes $\mathcal{R}_m$ into the ordinary scalar curvature supermultiplet $\mathcal{R}\equiv \mathcal{R}_0$, eqn.\eqref{curvature}. Thus $\mathcal{R}_m$ is, roughly speaking, the susy completion of the scalar curvature as redefined by the conformal/axial transformation given by multiplication by the first component of $T^m$;
that is, it is the susy completion of an expression of the form $e^{-2 f_m} (R-6 \partial_\mu f_m \partial^\mu f_m)$ for a certain function $f_m$ ($R$ being the bosonic scalar curvature). Of course $f_0=0$.
$\mathcal{R}_G$ may then be seen as the supermultiplet which gives the appropriate susy completion of, roughly,
\be
\sum_{n,m} n\, a_{n,m} ({\rm Re}\, T^{n+m-1}) e^{-2 f_m} (R-6\, \partial_\mu f_m \partial^\mu f_m).
\ee

Thus the difference between the ordinary curvature multiplet $\mathcal{R}_0$ and $\mathcal{R}_G$ is that the second one is the susy completion not just of the scalar curvature $R$ but rather of a generalized scalar curvature ($\psi R+ g_{T\bar T}\partial^\mu T \partial_\mu \bar T$) for some
$g_{T\bar T}$ and some function $\psi$ of the scalars.
For instance, if $H$ is quadratic in $C$, \textit{i.e.}\! the Lagrangian contains terms $[C \bar C]_D +....$, and $\partial_T F=const$, the resulting supergravity is, roughly speaking, the susy completion of
\be
R+ k_{T\bar T} \partial_\mu T\partial^\mu \bar T +(\psi R+ g_{T\bar T}\partial^\mu T \partial_\mu \bar T)^2
\ee
for some $K_{T\bar T}$, $\phi$ and $g_{T\bar T}$. This is what can be qualified as  higher derivative \textsc{sugra} coupled to $T$. Note that, by a conformal rescaling,
we may always reduce to the (susy completion of the) simpler form
\be
\phi R + K_{T\bar T} \partial_\mu T\partial^\mu \bar T +\rho R^2,
\ee
for some functions $\phi$, $\rho$ of the scalars and some \K,\ metric $K_{TT*}$. The dependence on $\alpha$  is hidden in these functions. Writing down the full component form of the Lagrangian is totally straightforward, but rather tedious.

\subsection{Generic $\alpha$--attractor models}

Generic $\alpha$--attractor models have the same duality properties as the $\alpha=1$ ones with stabilization terms
as in sec.\;\ref{inTT}, namely, they are dual to 4-derivative \textsc{sugra}'s with a local Lagrangian containing the \textit{same} number of chiral superfields. 

The generic $\alpha$--attractor models belong to the class of $N=1$ supergravities of the form\footnote{ Of course, we have a second class of allowed superpotentials having the  form $II)$ of eqn.\eqref{lagrangian2}.}
\begin{equation}\label{lastclass}
-\Big[\bar X^0X^0\Phi(T,\bar T, C, \bar C, Z_i, \bar Z_i)\Big]_D
+\Big(\Big[(X^0)^3\Big(C F(T, Z_i)+L(T,Z_i)+V(C,Z_i)\Big) \Big]_F+h.c.\Big)
\end{equation}
for a suitable choice of the arbitrary functions $\Phi$, $F$, $L$, $V$ ($\Phi$ being real, $F$, $L$, $V$ holomorphic).
For instance, the model \eqref{alpha} corresponds to no extra chiral fields $Z_i$ and 
\begin{gather}
\Phi(T,\bar T, C, \bar C)= \left(T+\bar T - C\bar C +3g\frac{(C\bar C)^2}{T+\bar T}\right)^{\!\alpha}\\
L(T)=V(C)=0.
\end{gather}
To show the duality, one goes through the same steps as in sec.\;\ref{inTT}. First one introduces two new $w=0$ superconformal chiral fields, $S$ and $Y$, and rewrites the Lagrangian in the form
\begin{multline}\label{longL}
-\Big[\bar X^0X^0\Phi(T,\bar T, Y, \bar Y, Z_i, \bar Z_i)\Big]_D
+\\
+\Big(\Big[(X^0)^3\Big(C F(T, Z_i)+L(T,Z_i)+V(Y,Z_i)+S(Y-C)\Big) \Big]_F+h.c.\Big).
\end{multline}
Again, integrating away the Lagrangian multiplier superfield 
$S$ we get $Y\equiv C$ reducing back to the model \eqref{lastclass}. The $T$ and $Y$ equations of motion give
\begin{align}
&\text{e.o.m. }\ T\colon &&\Sigma\big(\bar X^0\partial_T\Phi\big)=(X^0)^2\Big(C\,\partial_TF+\partial_TL\Big)\\
&\text{e.o.m. }\ Y\colon &&\Sigma\big(\bar X^0 \partial_Y\Phi\big)=(X^0)^2\, S,
\end{align}
which may be solved for $C$ and $S$, respectively
\begin{align}
&C=(X^0)^{-2}\, \frac{1}{\partial_T F}\,\Sigma\big(\bar X^0\partial_T\Phi\big)-\frac{\partial_TL}{\partial_TF}, && S=(X^0)^{-2}\, \Sigma\big(\bar X^0\partial_Y\Phi\big).
\end{align}
Plugging these expressions back into eqn.\eqref{longL}
we get the desired dual 4--derivative supergravity.

Let us specialize to the $\alpha$--attractor models where $L=V=0$. Then the Lagrangian is the sum of an ordinary (\textit{i.e.}\! 2--derivative) term $L_\text{ordinary}$ plus a higher derivative coupling $L_\text{4--der.}$. One has
\begin{gather}
L_\text{ordinary}= -\left[\bar X^0X^0\Big(\Phi-\frac{F}{\partial_TF} \,\partial_T\Phi -\frac{\overline{F}}{\overline{\partial_TF}}\, \partial_{\bar T}\Phi-Y\partial_Y\Phi-\bar Y \partial_{\bar Y}\Phi\Big)\right]_D\\
L_\text{4--der.}=-\left[\frac{\partial_T\Phi}{\partial_TF}\,\bar X^0\,\mathcal{R}^{(Y)}+\frac{\partial_{\bar T}\Phi}{\overline{\partial_TF}}\,X^0\,\overline{\mathcal{R}}^{(Y)} \right]_D
\end{gather}
where the curvature chiral multiplet  $\mathcal{R}^{(Y)}$ is defined as
\begin{equation}
\mathcal{R}^{(Y)}= (X^0)^{-1}\, \Sigma\big(\bar X^0 \partial_Y\Phi\big).
\end{equation}

\section{Discussion}

Current  cosmological observations have tested models of inflation in supergravity. One of the elegant models in the bosonic case is  Starobinsky $R+R^2$ model  \cite{Starobinsky:1980te} which was shown to be equivalent at the classical level to a model with one scalar field in \cite{Whitt:1984pd}. In supergravity the corresponding
 dual relation between the   2-derivative supergravity   interacting with two chiral matter multiplets  and pure higher curvature 4-derivative models without matter, was established  in  \cite{Cecotti:1987sa}. It was shown there that the duality transformation to a higher derivative model allows to eliminate both chiral matter multiplets. An elegant higher curvature pure supergravity model remains after duality transformation.
 
 In this note we have established the corresponding duality relation for a large class of more general cosmological models based on superconformal/supergravity models.  We have found that  it is possible to establish  a dual relation between  the   2-derivative supergravity   interacting with two chiral matter multiplets and the model with higher  4-derivative curvatures. However, in all cases, besides the so-called  $\alpha=1$ attractors models with a particular choice of the function $F(T)$ in the superpotential, corresponding to \cite{Starobinsky:1980te}, and with  specific stabilization, it is not possible to avoid interaction with scalars.
 The more we go away from the 1987 model \cite{Cecotti:1987sa}, the less elegant (and interesting) is the $R^2$ dual supergravity. However, ignoring the issue of  elegance, the higher derivative models presented above are  perfectly sound (and ordinary) $R^2$ supergravities interacting with scalars.  
  
 In expectation of the future cosmological data one may try to interpret our findings as follows.  First, consider the $\alpha=1$ attractors with the potential $f^2\big(\tanh {\varphi\over\sqrt{6}}\big)$ \cite{Kallosh:2013hoa}, including the one in $ \big(1- e^{-\sqrt {2\over 3 } \varphi} \big)^2$,
 \cite{Starobinsky:1980te}. We have found the dual higher supergravity underlying  all these models: all models but  the one corresponding to \cite{Starobinsky:1980te}  have higher curvature supergravities interacting with a scalar. Therefore for all of these models there is no simplification or elegance acquired from involving higher powers of the curvature superfield. If the B-modes will be detected at $r\approx 3\cdot 10^{-3}$ it will give a support to all models with $f^2\big(\tanh {\varphi\over\sqrt{6}}\big)$ \cite{Kallosh:2013hoa}. But  only one of them corresponds to a pure higher curvature supergravity, all others still have scalars.
 If  the B-modes are detected at  $r> 3\cdot 10^{-3}$,  we will have to switch to  $\alpha \neq 1$ superconformal/supergravity cosmological models  \cite{Kallosh:2013yoa}, which interpolate between chaotic inflation models \cite{Linde:1983gd} at large $\alpha$ and \cite{Starobinsky:1980te} at $\alpha=1$. The corresponding supergravities with higher curvature always have some interactions with scalars at $\alpha \neq 1$.  In this case, the idea that the inflaton is not a fundamental scalar but rather an excitation in a higher derivative gravity will be ruled out. In view of the recent discovery of the scalar Higgs particle, it will not be the first example  of the existence of scalar fields in nature. 
 
 Thus, the level of primordial gravity waves from inflation will clarify the relation  between models of ordinary supergravity with the inflaton as a matter multiplet and a possibility to replace it by a higher derivative supergravity with or without scalars. In the context of supersymmetric models described above we will learn from the B-modes if the inflaton is a fundamental scalar superfield or not.

\section*{Acknowledgements}

We acknowledge stimulating discussions with S. Ferrara, A. Linde,  D. Roest and A. Van Proeyen. RK is supported by the SITP and by the NSF Grant PHY-1316699 and by the Templeton foundation grant `Quantum Gravity Frontiers'.

\end{document}